\documentclass[twocolumn]{IEEEtran} 
\usepackage{epsfig}

\def\BibTeX{{\rm B\kern-.05em{\sc i\kern-.025em b}\kern-.08em
    T\kern-.1667em\lower.7ex\hbox{E}\kern-.125emX}}

\begin{document}

\title{High Temperature ${\pi \over 2}-$SQUID} 

\author{M.H.S. Amin, M. Coury, and G. Rose \thanks{All the authors are
associated with D-Wave Systems Inc., \#320-1985 West Broadway,
Vancouver, BC V6J 4Y3, Canada.  This work has been submitted to
the IEEE for possible publication.  Copyright may be transferred
without notice, after which this version may no longer be
accessible. }}


\maketitle

\begin{abstract}
A new DC-SQUID is proposed that exploits the properties of the
grain boundary junctions in high $T_c$ superconductors. The
orientations of the grain boundaries are chosen in such a way to
establish a $\pi/2$ (rather than 0 or $\pi$) phase difference
between the equilibrium phases of the two Josephson junctions in
the SQUID loop. This property is exploited to increase the
sensitivity and direction dependence of the SQUID for measuring
magnetic flux without additional flux generating coils.
\end{abstract}

\begin{keywords}
SQUID, High $T_c$ superconductivity, Quasiclassical theory.
\end{keywords}

\section{Introduction}

\PARstart{I}{n} a usual Josephson tunnel junction, the current $I$
passing through the junction is related to the phase difference
$\varphi$ between the two superconductors by \cite{tinkham}
\begin{equation}
I = I_{c} \sin \varphi. \label{I1}
\end{equation}
The equilibrium phase difference is zero in the absence of
current (current is also zero when $\varphi=\pi$, but this
corresponds to the maximum of the Josephson energy rather than a
minimum). The simple relation (\ref{I1}) should be replaced with
a more complicated one \cite{ko}, \cite{ko2} in the case of
constriction junctions, but the equilibrium phase difference will
still be at $\varphi=0$.

A conventional DC-SQUID consists of a superconducting ring with
two ordinary Josephson junctions on its opposite arms. The total
current passing through the SQUID is the sum of the currents
crossing each of the Josephson junctions. As a result of
interference between the two currents, the total current-phase
relation depends on the flux $\Phi$ threading the ring. In
particular, the critical current (the maximum current allowed
before going to non-stationary state with finite voltage drop
across the junction) is given by \cite{tinkham}
\begin{equation}
I_c = I_{c0} \left|\ \cos \left( {\phi_e \over 2} \right) \right|,
\label{Ic0}
\end{equation}
where $\phi_e$ is related to the external flux $\Phi$ threading
the loop through $\phi_e \equiv 2\pi \Phi/\Phi_0$, with $\Phi_0=
h/2e$ being the flux quantum, $h$ the Planck constant, and $e$
the charge of electron. Here we neglect the self-inductance of
the loop. Notice that $\phi_e$ in Eq. (\ref{Ic0}) has period
$2\pi$ which corresponds to one flux quantum. At small $\phi_e$,
the relation (\ref{Ic0}) is almost flat ($I_c \propto \phi_e^2$);
consequently, the sensitivity of the measurement of small fluxes
is low. Moreover, the critical current does not depend on the
direction of the magnetic field. To avoid this problem, one can
apply a biasing flux to move the working point of the SQUID away
from the flat region. However, this biasing flux increases the
noise as well as the complexity of the device and sometimes can
have unwanted influence on the measured system.

Therefore, it is desirable to shift the working point of the SQUID
away from the flat region without applying an external magnetic
field. This is possible by replacing one of the Josephson
junctions with a junction with non-zero equilibrium phase
difference. A DC-SQUID with a $\pi$-phase difference between the
equilibrium phases of the two Josephson junctions has already
been studied \cite{piSQUID}. Here we propose another type of
SQUID (we call it a $\pi/2$-SQUID) with the equilibrium phase
difference of $\pi/2$ between the junctions. The frustration
caused by the two junctions produces the desired shift of the
working point. In the next section we introduce such a device
using high $T_c$ superconducting grain boundary junctions. In
section III we study the $\pi/2$-SQUID using a simple model.
Section IV is devoted to the quasiclassical calculation of the
current in the $d$-wave structure and demonstrating agreement with
the simple model of section II. Finally, we summarize our results
in section V.

\section{High $T_c$ $\pi/2$-SQUID}

\begin{figure}[t]
\epsfysize 5cm \epsfbox[0 200 400 600]{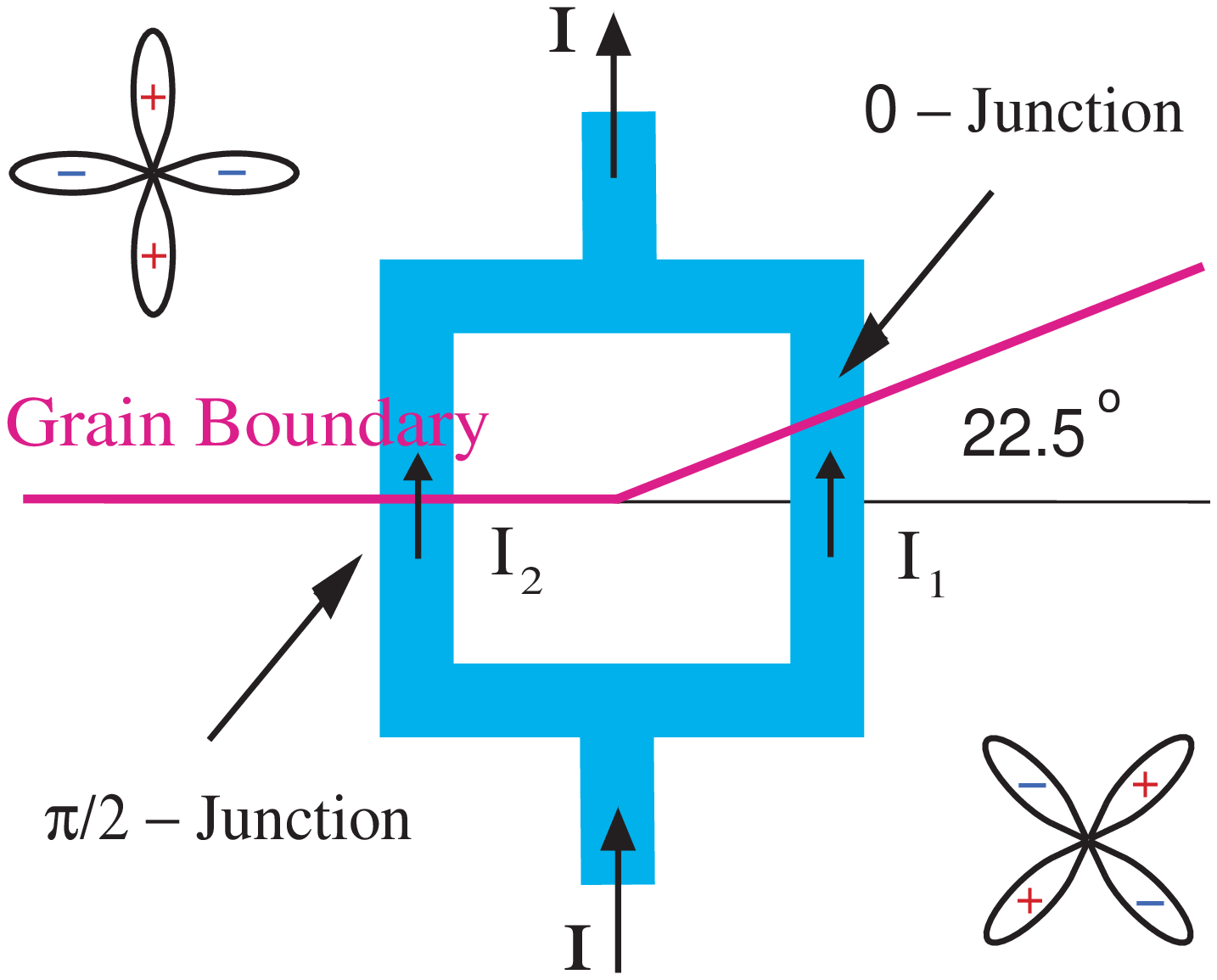}
\caption[]{\small Symmetric and Asymmetric $d$-wave grain boundary
junctions, with $45^\circ$ misorientation of the order parameters
on both sides, are used as two Josephson junctions of the
DC-SQUID.} \label{fig1}
\end{figure}

In high $T_c$ superconductors, because of the $d$-wave symmetry of
the order parameter, the current-phase relations are nontrivial
and generally depend on the orientations of the order parameters
on both sides of the junction \cite{gb}, \cite{gb2}. If we keep
only the first two harmonics, we can write the current-phase
relation as
\begin{equation}
I = I_1 \sin \varphi - I_2 \sin 2 \varphi. \label{I12}
\end{equation}
When $I_2 > I_1/2$ the equilibrium phase will occur at
$\varphi=\varphi_0$, where $\varphi_0$ is neither 0 nor $\pi$. In
particular, when the first harmonic vanishes, the equilibrium
phase difference happens at $\varphi_0=\pi/2$. These types of
junctions, known as $\pi/2$-junctions, can be realized at the
grain boundary junction between two $d$-wave superconductors with
crystalline orientations of $0^\circ$ and $45^\circ$ with respect
to the grain boundary. At such a junction (known as an asymmetric
junction), there is an exact cancellation of the first harmonic
and the second harmonic dominates the current \cite{gb},
\cite{gb2}. These junctions have been realized and tested
experimentally \cite{ilichev1}. A symmetric junction can also be
made with $45^\circ$ misorientation angle by choosing
-22.5$^\circ$ and $22.5^\circ$ crystalline orientations at the
two sides. This junction is very different from the asymmetric
one; specifically, the exact cancellation of the first harmonic
does not occur in this case \cite{gb2}. In a realistic junction
with non-ideal transparency and roughness at the boundary, the
second harmonic gets suppressed by the boundary imperfections more
strongly than the first harmonic and, as a result, the latter
dominates the current. This leads to a conventional junction
(0-junction) with the usual current-phase relation of Eq.
(\ref{I1}). These junctions have also been studied experimentally
\cite{ilichev2}.

By combining symmetric and asymmetric junctions, we can have a
SQUID with $\pi/2$ equilibrium phase difference between the
junctions. Fig.\ 1 illustrates such an structure. First we study
this system using a simple model. The more realistic numerical
simulation of the system using a quasiclassical model will come
after that.

\section{Simple Model}

Consider a DC-SQUID, consisting of a conventional and a $\pi/2$
Josephson junction, with current-phase relations
\begin{equation}
I_1 = I_{c1} \sin \varphi_1, \qquad I_2 = - I_{c2} \sin 2
\varphi_2, \label{cp}
\end{equation}
respectively. To simplify the calculations, we will assume
$I_{c1}=2I_{c2}=2I_{c0}$.  The more general case of arbitrary
$I_{c1}/I_{c2}$ can also be treated as the solutions are similar.
We also neglect the self inductance of the ring in our
calculations. When there is a flux $\Phi$ threading the SQUID
ring, the phases at two junctions are related by $\varphi_1 -
\varphi_2 = \phi_e$. The total current passing through the SQUID
is then given by
\begin{equation}
I = I_{c0} [ 2\sin (\varphi_2 + \phi_e) -  \sin 2 \varphi_2].
\label{cpr}
\end{equation}
The critical current is the maximum current in this current-phase
relation. To find the critical current, one has to find the
solutions to the equation $d I/d \varphi_2=0$. These solutions are
$\phi_e+2\pi n$ and $(-\phi_e + 2\pi n)/3$, with $n$ being an
integer number. Substituting into Eq. (\ref{cpr}) we find the
maximum current to be

\begin{equation}
I_c = {3 \over 2}\ I_{c0}\ {\rm Max} \left\{ \sin {2 \over
3}(\phi_e + n \pi) \right\}. \label{Ic}
\end{equation}
Here ``Max'' means the maximum of the three cases with $n=0,\pm
1$.

\begin{figure}[t]
\epsfysize 5cm \epsfbox[70 280 400 550]{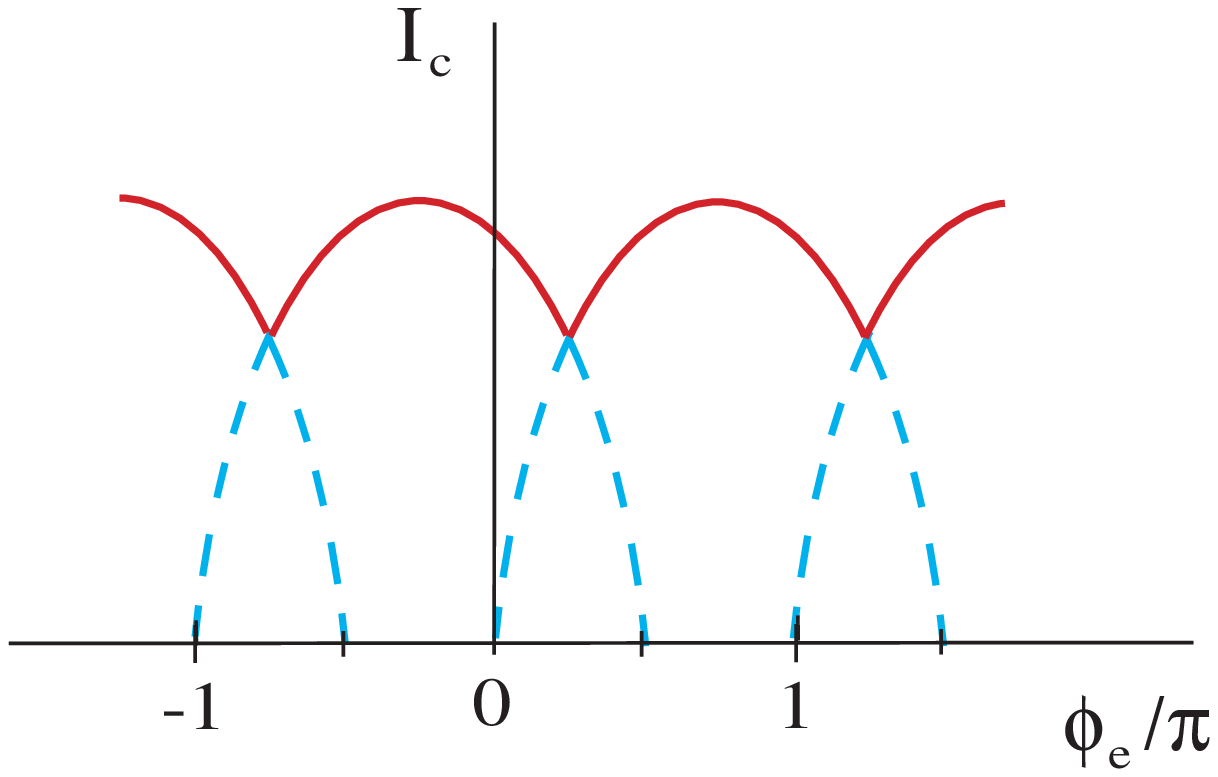} \caption[] {
Solid line is the critical current as a function of external flux
for the simplified model. Dashed line shows the sine-functions in
Eq. (6).} \label{fts}
\end{figure}

Fig.\ 2 shows the flux dependence of the critical current using
Eq.\ (\ref{Ic}). Notice that the periodicity of the function is
$\pi$ (half a flux quantum) although the period of the argument of
the sine-functions in Eq.\ (\ref{Ic}) is actually $3\pi$. The
$\pi$-periodicity of Eq.\ (\ref{Ic}) is a result of the symmetry
$I(\phi_e+\pi,\varphi_2)= I(\phi_e,\varphi_2+\pi)$ in Eq.\
(\ref{cpr}). In other words, changing the applied flux by half a
flux quantum shifts the current-phase relation by $\pi$ and
therefore does not change the critical current. This is evidently
different from the ordinary SQUID for which the period is $2\pi$
(one flux quantum). The important feature here is that the
$\phi_e=0$ does not coincide with the maximum of the function as
it does in usual the DC-SQUIDs. Therefore, an increase in
sensitivity for measuring small fluxes is gained. Moreover, the
direction of the flux is also detectable by measuring the
critical current.

\section{Quasiclassical Calculations}

As was mentioned, the current-phase relations in $d$-wave grain
boundary junctions does not usually have the simple forms of Eq.\
(\ref{cp}). One has to find the realistic current-phase relation
using a microscopic theory, taking into account the imperfections
of the boundary. Here we use a quasiclassical model to study this
system. The equations we solve are the Eilenberger equations
\cite{eilenberger}

\begin{equation}
{\bf v}_{F}\cdot \frac{\partial }{\partial {\bf
r}}\widehat{G}_{\omega }({\bf v}_{F},{\bf r})+[\omega
\widehat{\tau }_{3}+\widehat{\Delta }({\bf v}_{F},{\bf
r}),\widehat{G}_{\omega }({\bf v}_{F},{\bf r})]=0,  \label{EqA1}
\end{equation}
where
\begin{equation}
\widehat{\Delta }=\left(
\begin{array}{cc}
0 & \Delta \\
\Delta ^{\dagger } & 0
\end{array}
\right), \quad \widehat{G}_{\omega }({\bf v}_{F},{\bf r})=\left(
\begin{array}{cc}
g_{\omega } & f_{\omega } \\
f_{\omega }^{\dagger } & -g_{\omega }
\end{array}
\right).
\end{equation}
$\Delta $ is the superconducting order parameter and
$\widehat{G}_{\omega }({\bf v}_{F},{\bf r})$ is the matrix Green's
function, which depends on the electron velocity on the Fermi
surface ${\bf v}_{F}$, the coordinate ${\bf r}$, and the Matsubara
frequency $\omega=(2n+1)\pi T$, with $n$ being an integer number
and $T$ the temperature. We also need to satisfy the
normalization condition
\begin{equation}
g_{\omega }=\sqrt{1-f_{\omega }f_{\omega }^{\dagger }}.
\label{EqA4}
\end{equation}
In general, $\Delta $ depends on the direction of ${\bf v} _{F}$
and is determined by the self-consistency equation

\begin{equation}
\Delta ({\bf v}_{F},{\bf r})=2\pi N(0)T\sum\limits_{\omega >0}
\left< V( {\bf v}_{F},{\bf v}_{F}^{\prime })f_{\omega }({\bf
v}_{F}^{\prime }, {\bf r}) \right>_{{\bf v}_{F}^{^{\prime }}}
\label{GapEq}
\end{equation}
where $V({\bf v}_{F},{\bf v}_{F}^{\prime })$ is the interaction
potential. Solution of matrix equation (\ref{EqA1}) together with
(\ref{GapEq}) determines the current density $ {\bf j(r)}$ in the
system

\begin{equation}
{\bf j(r)}=-4\pi ieN(0)T\sum\limits_{\omega >0} \left<{\bf
v}_{F}g_{\omega }({\bf v}_{F},{\bf r}) \right>_{{\bf v}_{F}}.
\label{EqA6}
\end{equation}
In two dimensions, $N(0)=m/ 2\pi$ is the 2D density of states and
$\left<... \right>=\int\limits_{0}^{2\pi } (d\theta / 2\pi )...$
is the averaging over directions of the 2D vector ${\bf v }_{F}$.

\begin{figure}[t]
\epsfysize 6cm \epsfbox[50 120 400 470]{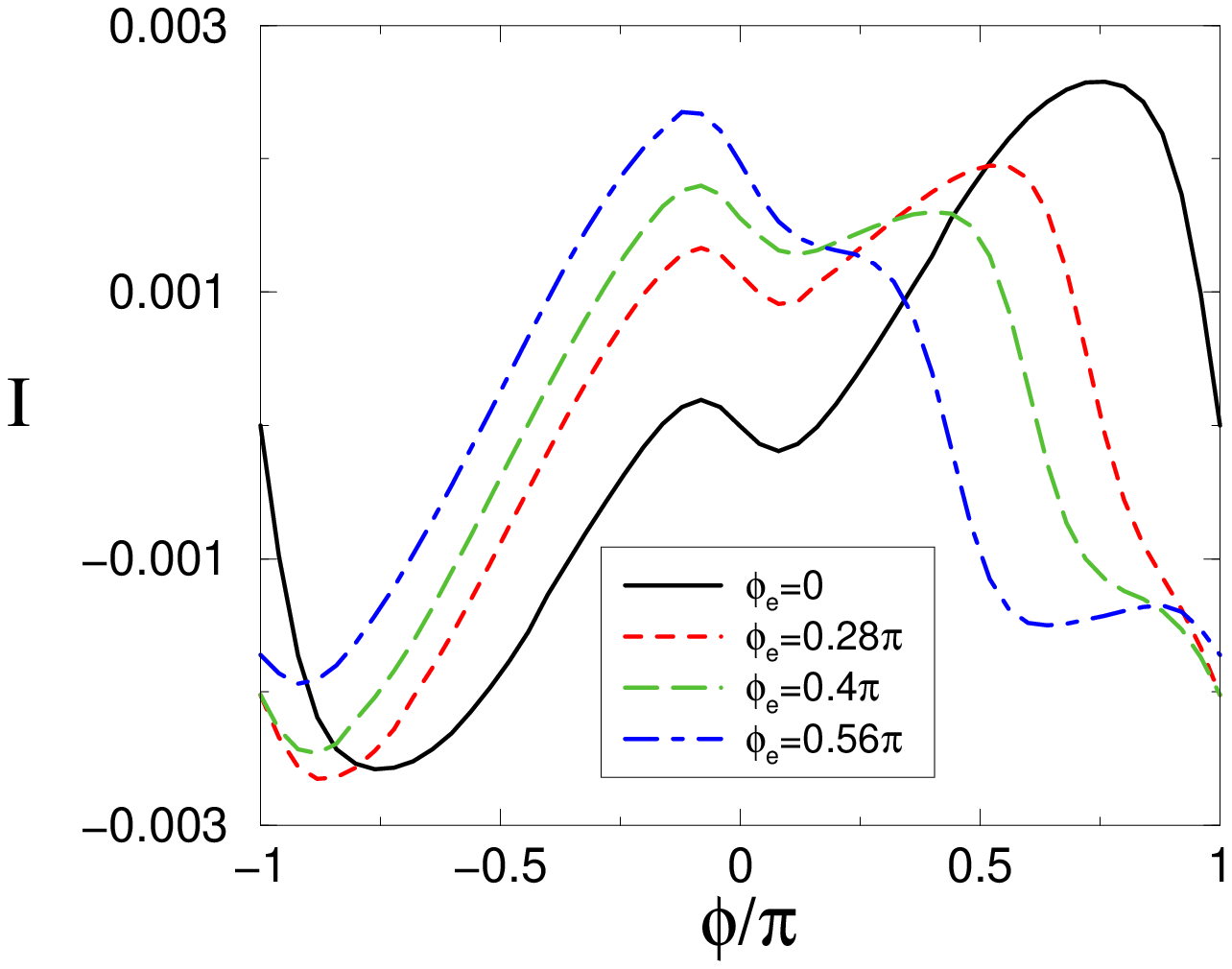}
\caption[]{Current-phase relation obtained from the
quasiclassical model using $r=1.4$.} \label{fig3}
\end{figure}

The transparency of the boundary is described by the parameter
$D_0$ \cite{amin1}, \cite{amin2}. To incorporate the roughness, we
assume a narrow scattering region of width $d$ between the two
superconductors \cite{amin1}, \cite{amin2}. The roughness is
therefore parameterized by $\rho = d/l$, where $l$ is the mean
free path of the quasiparticles in this scattering region. In our
calculations, we use $D_0=\rho = 0.5$ which are reasonable values
for a realistic system. Different widths, $W_1$ and $W_2$, are
chosen for the two arms and the ratio of the widths is denoted by
$r=W_2/W_1$. The details of our numerical method for solving
these equations is given in \cite{amin2}.

Fig.\ 3 displays the current-phase relation at different values
of the external flux $\phi_e$ for a SQUID structure with $r=1.4$.
At $\phi_e=0$, the graph has two local maxima, which is the
signature of having a mixture of first and second harmonics. As
$\phi_e$ is increased, the two maxima become closer and at some
value of $\phi_e$, they will be equal. At this point, the global
maximum jumps from one local maximum to another. This actually
explains the sharp kinks at the minimum positions in Fig.\ 2 (and
also in Fig.\ 4).


Fig.\ 4 shows the critical current as a function of $\phi_e$ for
different values of $r$. One can immediately see the similarities
between these graphs and the one obtained from our simple model
in Fig.\ 2. Specifically, the maxima again occur with period of
$\pi$, although the $\pi$-periodicity is not exact here (the real
period is $2\pi$). The maximum of the critical current is shifted
from $\phi_e=0$, as was desired. One can also see from Fig.\ 4
that the maximum range of variation of $I_c$ is obtained at
$r=1.4$. This is actually the optimal ratio of the widths for
SQUID design.

\begin{figure}[t]
\epsfysize 6cm \epsfbox[50 120 400 470]{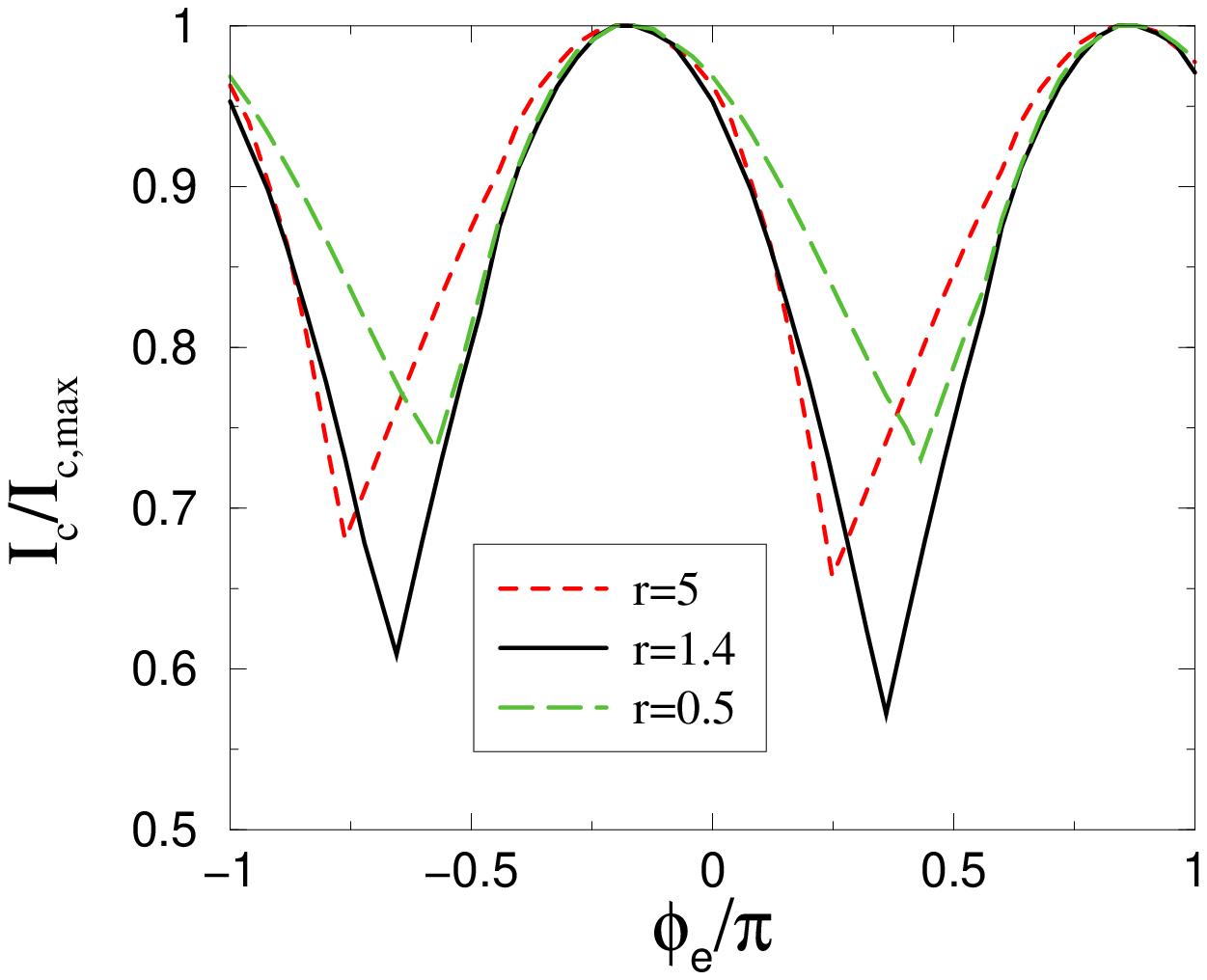}
\caption[]{Critical current versus the external flux for
different values of $r$. } \label{fig4}
\end{figure}

\section{Conclusions}

We proposed and studied a DC-SQUID which has $\pi/2$ difference
between the equilibrium phases at its two junctions. The $\pi/2$
phase difference is easily obtained using the properties of high
$T_c$ grain boundary symmetric and asymmetric junctions. The
frustration caused by this phase difference moves the working
point of the SQUID to a point at which the maximum critical
current is obtained at a non-zero external flux. This will
increase the sensitivity of the SQUID at small magnetic field,
with a dependence on the direction of the flux, without any
external biasing coil.

We introduced a simple model and compared its results with the
ones obtained from a microscopic quasiclassical model. The
agreement between the two models is acceptable. We found that the
$\pi/2$-SQUID exhibits other new features, different from a
conventional SQUID or a $\pi$-SQUID, among those is the $\Phi_0/2$
(instead of $\Phi_0$) periodicity of the critical current. We
find the optimal ratio between the widths of the two arms to be
$r=1.4$.

\section*{Acknowledgments}

The authors are grateful to A.M. Zagoskin, E. Il'ichev, and J.P.
Hilton for stimulating discussions.

\nocite{*}
\bibliographystyle{IEEE}

%

\end{document}